\documentclass{article}

\usepackage{arxiv}

\usepackage[utf8]{inputenc} 
\usepackage[T1]{fontenc}    
\usepackage{hyperref}       
\usepackage{url}            
\usepackage{booktabs}       
\usepackage{amsfonts}       
\usepackage{nicefrac}       
\usepackage{microtype}      
\usepackage{graphicx}
\usepackage{doi}
\usepackage{blindtext}
\usepackage{enumitem}
\usepackage{epsfig}
\usepackage{graphicx}
\usepackage{amsmath}
\usepackage{amssymb}
\usepackage{multicol}
\usepackage{float}
\usepackage{mcite}
\usepackage[euler]{textgreek}
\usepackage{subcaption}

\setitemize[0]{leftmargin=*}

\renewcommand{\headeright}{}
\renewcommand{\undertitle}{}
\renewcommand{\shorttitle}{"Partially Wet Phase" MCB-Junction}

\hypersetup{
pdftitle={},
pdfsubject={},
pdfauthor={},
pdfkeywords={},
}

\begin{document}
\title{A pivoting Mechanically Controllable Break junction setup enabling “partially wet phase” MCB-junctions.}

\author{C.J. Muller\thanks{Correspondence: PO Box 60, 6580 AB Malden, The Netherlands}
\\ \\
This research has, in its entirety, been privately conducted and funded by the author.
}
\maketitle

\begin{abstract}
A technique is presented, which creates MCB junctions that can be pivoted to any desirable angle. The MCB junction equipped with a specific glass liquid cell can be used to produce a MCB junction, of which the electrodes are covered with a microscopic layer of fluid, thus producing a “partially wet phase” MCB junction.
\end{abstract}
\begin{multicols}{2}

\section{Introduction}
The development of the point-contact-spectroscopy technique by gently pressing a needle on a metallic bulk piece dates back to 1966 \cite{levinstein1966source0}. The so called “spear anvil” technique has been very successful in studying the density of states and the superconducting gap at the time. The contacts were usually created by a sharp screw of a certain metal which was gently contacting the anvil by turning the screw. At the microscopic scale the contacts were not well defined, both tunnel junctions created by the native metal-oxide as well as multiple micro metallic contacts could coexist in the same contact. Nowadays the point contact technique has evolved to provide scientists with an expanded toolset including STM versatility and MCB junctions. 

The MCB technique \mcite{muller1992source1a,*muller1992source1b,*muller1996source1c} has been used to obtain insights in a wide spectrum of physical science. For example, molecules can be captured between the atomically sharp electrodes, enabling a host of experiments on a single captured molecule. The electrodes themselves can be made up from normal metals, superconductive metals or exotic materials, for example heavy fermion superconductors. The adjustable contact size of the constriction at the atomic scale, the possibility to move atoms in and out of this constriction in a reproducible way, as well as the adjustable vacuum gap in the non-contact mode ensures a flexible baseline setup. Experiments can be performed in vacuum (from dilution fridge temperatures to room temperature), or at ambient conditions, or submerged in a cryogenic fluid. These possibilities have led to a wealth of discoveries made in this interesting field, which basically studies matter at the atomic and/or molecular scale. 

\begin{figure}[H]
    \centering
    \includegraphics[width=0.4\textwidth]{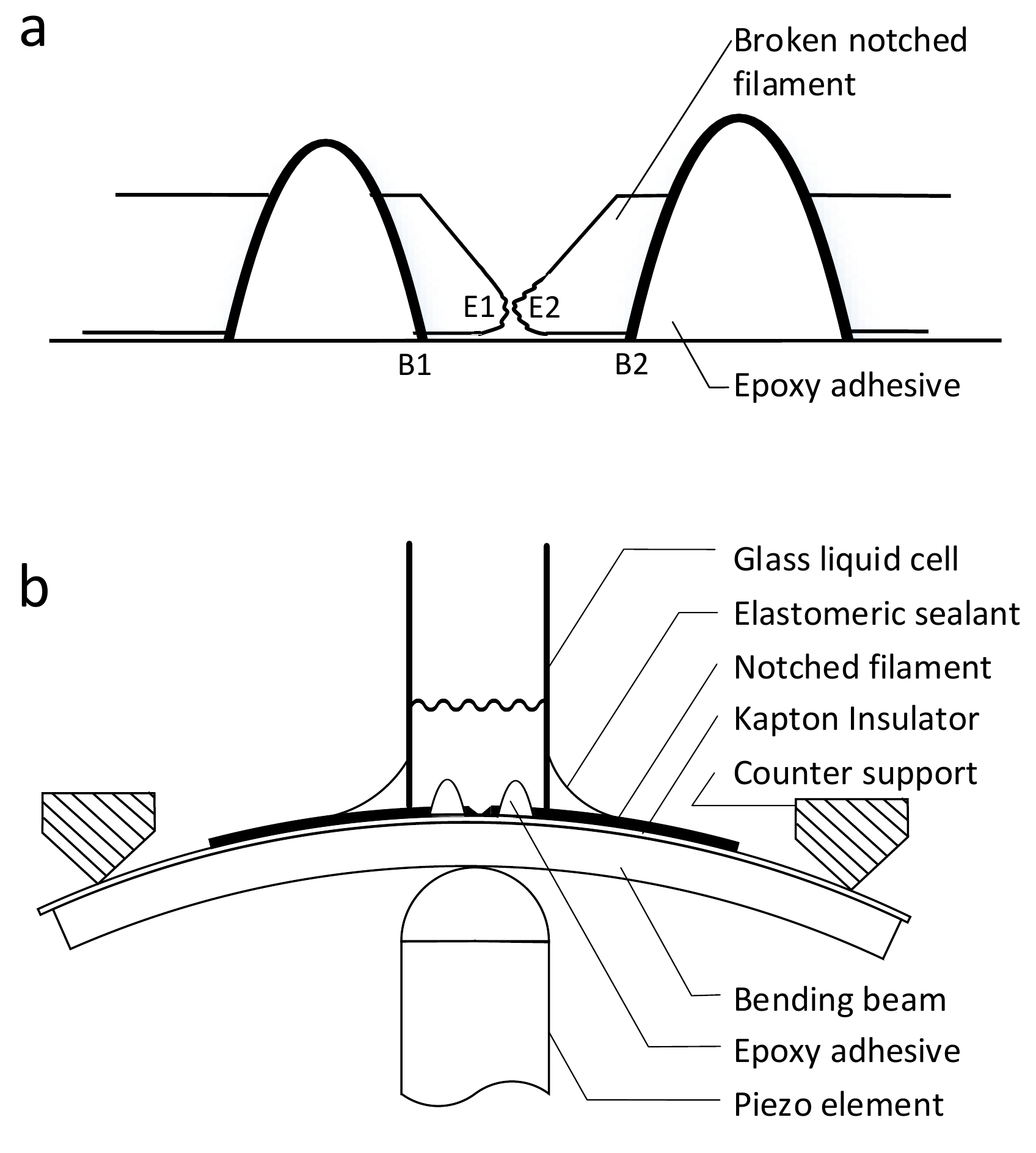}
    \caption{Figure 1a shows a schematic of the central section of a notched MCB junction. The small size of the electrode-bending beam loop, electrode1-bendingbeam1-bendingbeam2-electrode2 (E1-B1-B2-E2) plays a key part in the MCB operation. Figure 1b shows a MCB liquid cell with the junction fully submerged in the liquid which has been used in molecular conductivity experiments \protect\cite{muller1992source1a}}
    \label{fig:figure1}
\end{figure}

The novelty of the MCB junction in the early 90’s relied on the heart of the device. The electrode-bending beam loop, E1-B1-B2-E2 as indicated in Figure \ref{fig:figure1}a, is on the order of 100 to a few hundred micrometers for notched MCB junctions. It is feasible to create a distance B1-B2 of 1 to 2 filament diameters, down to 25 micrometer filaments. For lithographically defined MCB junctions this loop can be as small as a micrometer to a few tens of micrometers. The small dimensions of the electrode-bending beam loop ensure an extreme rigidity of the system, not susceptible to all kinds of (building) vibrations, present in most labs. A large attenuation factor, which is inversely proportional to the electrode-bending beam loop, exists between the extension of the actuator perpendicular to the bending beam, as compared to the movement of the electrodes alongside the bending beam. Making the electrode-bending beam loop larger, millimeters or centimeters, will facilitate breaking of the electrode material. However this will come at the cost of reduced rigidity and stability as well as a reduction of the attenuation factor. This will thus in general not lead to the subatomic adjustability and stability that has become the hallmark of MCB junctions.

One specific category of MCB experiments is submerged in a fluid at ambient conditions. Some special precaution is required to ensure operability. For example, one opening of a small glass capillary can be glued on the bending beam with a flexible elastomeric sealant surrounding the junction area, see Figure \ref{fig:figure1}b. If the capillary is positioned vertically, a liquid can be injected at the other end to ensure the MCB junction is submerged. This method has been proven \cite{muller1992source1a}, i.e. the junction can be broken and remains mechanically controllable whilst being submerged in a liquid. The flexible elastomeric sealant ensures a tight seal; it caters as well for various degrees of flexing of the MCB’s bending beam. A controllable junction submerged in a fluid opens a new range of physical science to be studied. It is possible to measure the effect of adhesion of a fluid to the electrode material, making it harder for the electrodes to contact when brought together, as opposed to a situation in vacuum. In addition specific molecules can be dissolved in the fluid. These molecules may be tailored with anchor groups to have strong covalent bonds to the electrodes either on one side, or on opposite sides and bridge the two junction electrodes once a matching gap is adjusted. In this way the molecular deposition is modifying the transport characteristics within the fluid environment.

\section{Pivoting MCB design}
\label{sec:Pivoting MCB design}

\begin{figure}[H]
    \centering
    \includegraphics[width=0.45\textwidth]{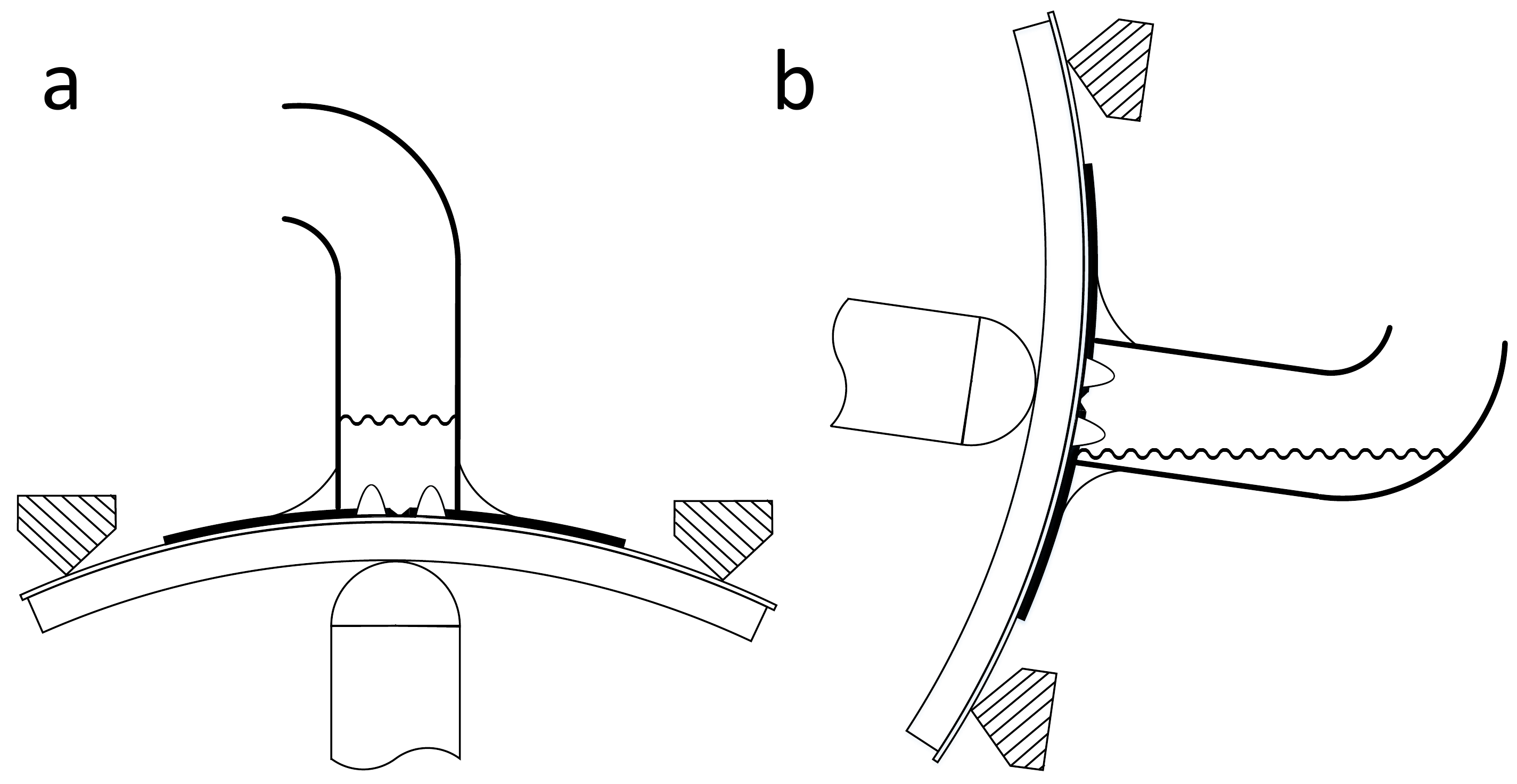}
    \caption{A MCB liquid cell as detailed in the text. When used in the pivoting MCB setup, it enables both the completely wet phase (Figure 2a) as well as the partially wet phase (Figure 2b). The partially wet phase depends on the time after draining the junction area, until a microscopic layer of fluid resides on the junction electrodes.}
    \label{fig:figure2}
\end{figure}

Here we will focus on a MCB design variant which is flexible to pivot an adjusted MCB junction, leaving the adjusted junction intact during pivoting. The pivoting MCB design operation is aimed at ambient conditions, where the setup can either submerge the MCB junction in a fluid, or drain the fluid from the junction area depending on a certain pivot angle. Its basic principle relies on the pivoting of a liquid cell, which holds the liquid. The cell is designed in such a manner that it can hold the liquid while being tilted at various angles. At one position the MCB junction is submerged in the liquid (see Figure \ref{fig:figure2}a). At another position the MCB junction is drained, while the cell still contains the fluid (see Figure \ref{fig:figure2}b).

\begin{figure}[H]
    \centering
    \includegraphics[width=0.45\textwidth]{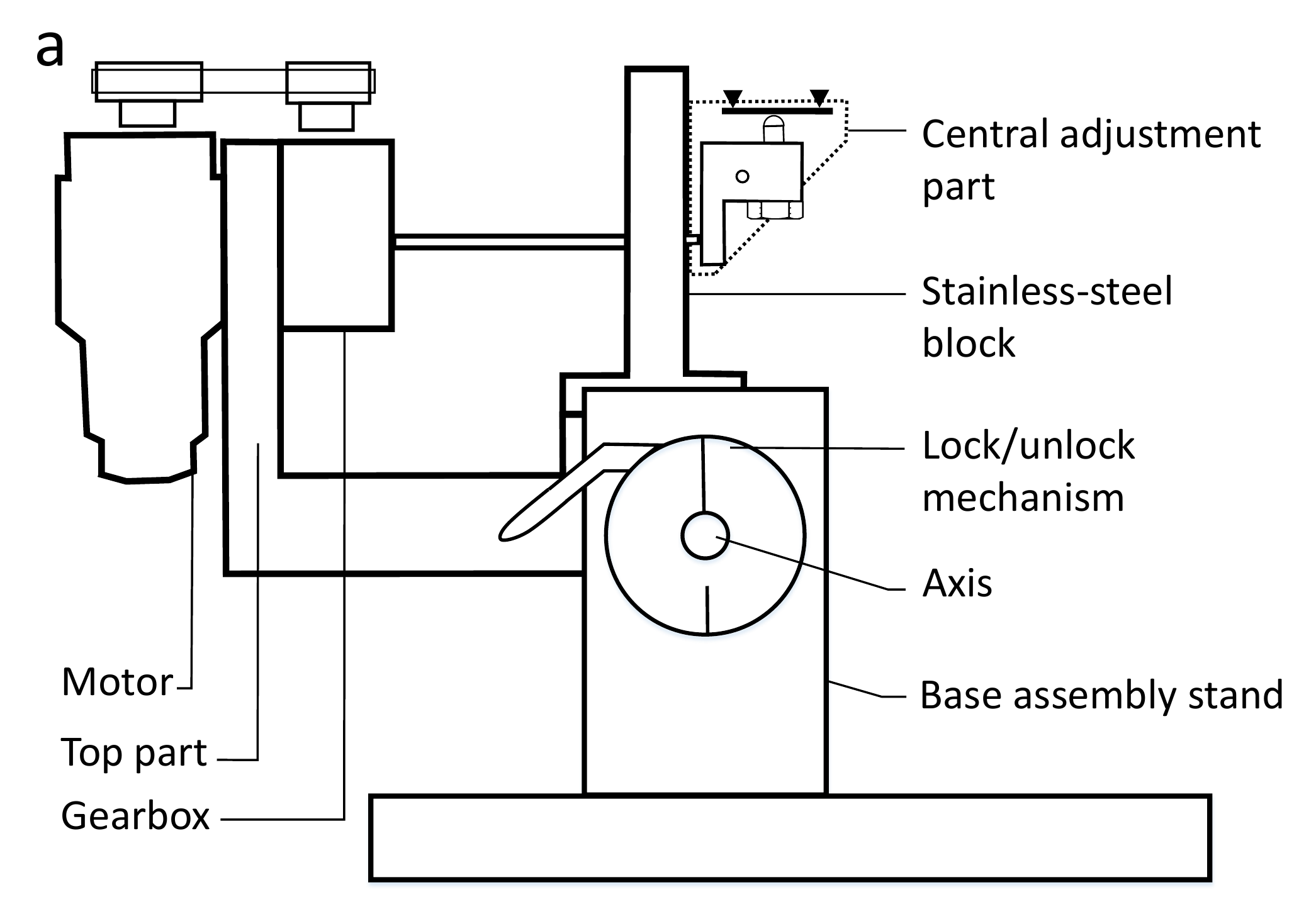}
\end{figure}
\begin{figure}[H]
    \centering
    \includegraphics[width=0.34\textwidth]{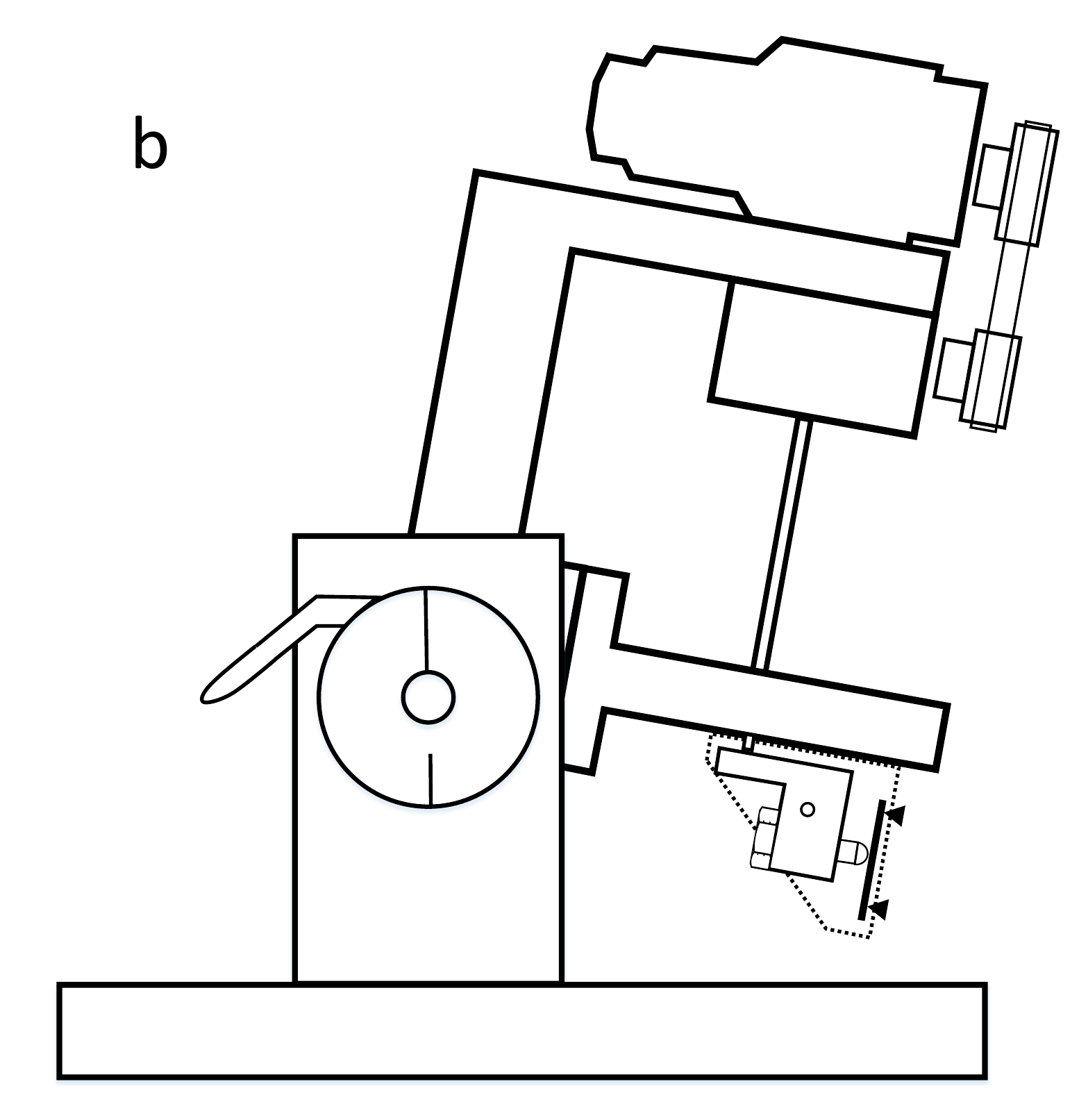}
    \caption{A schematic view of the pivoting MCB setup. The MCB pivot angle in figure 3a and 3b correspond to the two sample positions indicated in figure 2a and 2b respectively.}
    \label{fig:figure3}
\end{figure}

This MCB setup encompasses three main parts bolted solidly together and mounted on a spindle which can freely rotate in a base assembly stand. The three main parts are: a classical central adjustment part, a top part and a solid stainless-steel block, connecting the prior two parts, see Figure \ref{fig:figure3}.

The MCB bending beam is flexed using a motor and gearbox 1:120, driving a spindle which is coupled to a screw. Depending on the revolution direction of the motor, this screw moves a piezo towards or away from the bending beam via a lever. For fine tuning of the MCB junction, a hard-piezo material can be used, extending 5\textmu m at 1000V. The central adjustment part contains the screw, the piezo, the lever as well as the bending beam assembly. The central adjustment part is mounted to a stainless-steel block, which forms the right leg of a rigid, solid “U-shaped” part (see Figure \ref{fig:figure3}a). The top part forms the left leg of this U-shape in Figure \ref{fig:figure3}a. 

\begin{figure}[H]
\centering
\begin{subfigure}{0.4\textwidth}
    \centering
    \includegraphics[width=\textwidth]{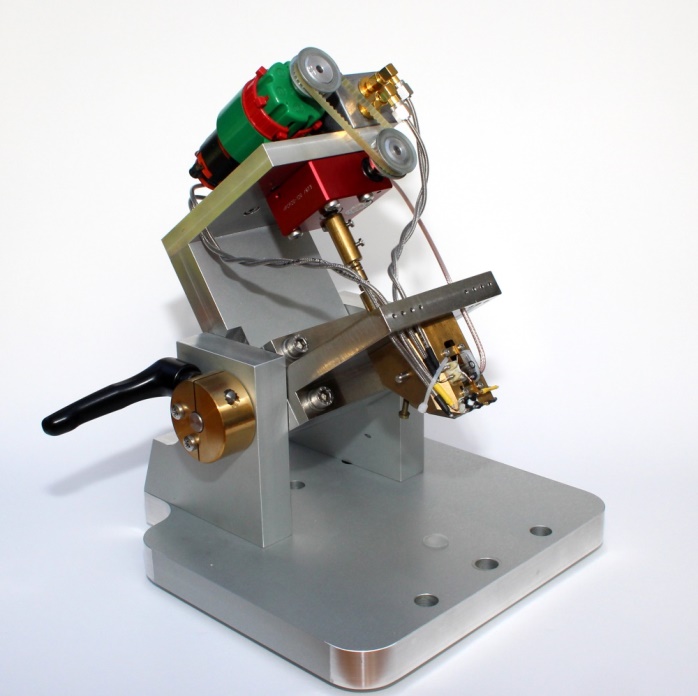}
\end{subfigure}
\begin{subfigure}{0.4\textwidth}
    \centering
    \includegraphics[width=\textwidth]{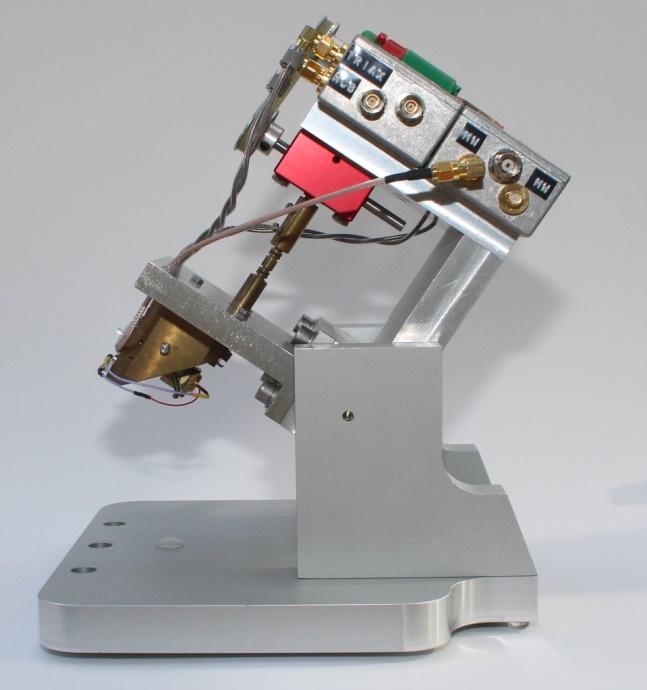}
\end{subfigure}
    \caption{Photos of the pivoting MCB setup from 2 different angles}
    \label{fig:figure4}
\end{figure}

The motor and gearbox are mounted to the top part. Eddy current metal shielding boxes are mounted to the top part. Electrical connections, running to the central adjustment part, enter these boxes via corresponding feedthrough plugs/connectors. In this way the adjusted MCB junction is not influenced by the connecting measurement cables. Instead these cables pull on the top part, where they do not influence the junction fixture, see Figure \ref{fig:figure4}.

The entire setup can be pivoted at the lower right corner of the “U”, around a spindle in a base assembly stand. It can be manually locked/unlocked in any desired position by a shaft collar with a clamping lever, see Figure \ref{fig:figure3}, \ref{fig:figure4}.

\section{Modes of operation}
A typical mode of operation would make use of the flexible pivot-angle design. For example, consider a pivot angle in which the liquid cell is in a vertical position (Figure \ref{fig:figure2}a). In this case an adjusted and fine-tuned MCB junction is submerged in fluid. By modifying the pivot-angle to arrive at a liquid cell in a near-horizontal position (Figure \ref{fig:figure2}b), the same adjusted and fine-tuned MCB junction is now drained from liquid. This procedure can be repeated, as the liquid is contained by the liquid cell at these pivoting angles. 

With this design, repeatability can be studied of both submerged junctions as well as drained junctions. In analogy to the physics of wetting \cite{bonn2009source2}, the following definitions of possible phases for the MCB junction are proposed:

\begin{itemize}
    \item Completely wet phase: A MCB junction submerged in a liquid, or if a macroscopic layer of liquid wets its electrodes.
    \item Completely dry phase: A MCB junction free from the liquid used in the experiment or drained from this liquid, and left long enough to dry, to be sure no liquid layer is present on the junction’s electrodes
    \item Partially wet phase: A MCB junction drained from its liquid, however, where the junction electrodes still hold on to a microscopic layer of liquid. A microscopic layer may consist of a thickness of one or several liquid molecules, depending on the liquid surface affinity.
\end{itemize}

Partially wet phase MCB junctions are in general difficult to obtain. Once drained, the junction electrodes areas will start with a macroscopic layer of liquid adhered to it, and thus show behavior as in the “Completely wet phase”. As time progresses, the junction electrodes hold less and less layers of fluid due to evaporation, until the electrodes will be completely dry from this fluid. So why try to measure on a system which is time dependent anyhow? The answer is manifold. 

First, at the atomic/molecular or mono-layer level, physics is different from our everyday experiences. At ambient conditions, all solid surfaces are wetted with some layer of fluid \cite{bonn2009source2}. If we are able to control the type of fluid, the resulting microscopic layer of fluid is sustainable over a relatively long time period. This will provide ample time to perform measurements on a fixed, unique and well defined system.

Second, within solid state physics we have come to appreciate the importance of interfaces on electrical transport parameters. A few examples of these phenomena are as follows:

\begin{itemize}
    \item A textbook example here is the PN junction interface. At and near the interface, the excess of charge carriers in the P material (positive holes) will need to rebalance with the excess of electrons in the N material. Over a characteristic distance extending from the interface into the P and the N region, called the depletion region, a build-in potential is generated by positive carriers in the N region and negative carriers in the P region. Thus establishing equilibrium between the generated electric field in the depletion region and the diffusion process because of the concentration differences in charge carriers. 
    \item The metal-semiconductor interface is another interesting one.  Here negative surface states at the metal surface and a positive charge distribution over a finite region in the semiconductor make up for the well-known Shottky barrier \cite{schottky1939source3}.
    \item With proper bandgap engineering the GaAs AlGaAs hetero junction can be manufactured such that at the GaAs/AlGaAs interface a “V-shaped” triangular potential exists just below the Fermi level, enabling filling one or a few levels within this potential trap with electrons. At low enough temperatures this leads to a two-dimensional electron gas (2DEG) at the GaAs/AlGaAs interface. 2DEG’s have played an instrumental role in the discovery of the quantum Hall effect \cite{klitzing1980source4} as well as in the understanding of conductance in mesoscopic structures (for a review see \cite{kouwenhoven1997source5}).
    \item Graphene is a more recent and extreme example. This material of a mono atomic layer of carbon atoms is interfaced above and below plane with a high barrier, disabling the layer to layer interaction which is present in graphite. Graphene reveals the quantum Hall effect at room temperature \cite{novoselov2007source6}, which is absent in graphite, which consists of many weakly coupled graphene layers.
\end{itemize}

Third, however thin a microscopic layer of fluid may be, it can provide structural support. The microscopic layer adheres strongly to the electrode surface, thus it may provide structural support for a bridging molecule by wrapping it in a microscopic 3-dimensional layer lattice.

Summarizing after a MCB junction is drained a macroscopic layer of liquid will wet its electrodes. If left untouched to dry, at some point the “Partially wet phase” will be entered. Here only a microscopic layer of fluid exists on the junction electrodes. This layer will evaporate eventually and be replaced by a layer of $\text{H}_2\text{O}$ at ambient conditions. Thus, although the electrodes may be completely dry from the specific fluid used in the experiments, a “Completely dry phase” will not be reached at ambient conditions.

\vfill\null
\columnbreak
\section{Conclusion}
We have provided a design of a pivoting MCB setup which enables studies of the partially wet phase MCB junction. Controlled “partially wet phase” MCB junctions have been difficult to obtain until now. These junctions may show interesting physics, as electrical properties as well as mechanical properties at the molecular level can be profoundly influenced by the presence of a microscopic layer of fluid.

\section*{Acknowledgement}
We would like to thank H.A.J. Peters for the construction of the pivoting MCB setup. 

\bibliographystyle{unsrt}
\bibliography{main}

\end{multicols}

\end{document}